\documentclass[fleqn,10pt]{wlscirep}
\usepackage[utf8]{inputenc}
\usepackage[T1]{fontenc}
\usepackage{mathrsfs}
\usepackage{lineno}
\usepackage{aas_macros}
\usepackage{soul}

\title{Oscillations of Highly Magnetized Non-rotating Neutron Stars}

\author[1,2,*]{Man Yin Leung}
\author[1]{Anson Ka Long {Yip}}
\author[1,3,4]{Patrick Chi-Kit {Cheong}}
\author[1,5,6]{Tjonnie Guang Feng {Li}}
\affil[1]{Department of Physics, The Chinese University of Hong Kong, Shatin, N. T., Hong Kong}
\affil[2]{Department of Physics, The Hong Kong University of Science and Technology, Clear Water Bay, Kowloon, Hong Kong}
\affil[3]{Department of Physics and Astronomy, University of New Hampshire, 9 Library Way, Durham NH 03824, USA}
\affil[4]{Department of Physics, University of California, Berkeley, Berkeley, CA 94720, USA}
\affil[5]{Institute for Theoretical Physics, KU Leuven, Celestijnenlaan 200D, B-3001 Leuven, Belgium}
\affil[6]{Department of Electrical Engineering (ESAT), KU Leuven, Kasteelpark Arenberg 10, B-3001 Leuven, Belgium}

\affil[*]{emilylmy1030@gmail.com}

\makeatletter
\renewcommand{\@maketitle}{%
{%
\thispagestyle{empty}%
\vskip-36pt%
{\raggedright\sffamily\bfseries\fontsize{20}{25}\selectfont \@title\par}%
\vskip10pt
{\raggedright\sffamily\fontsize{12}{16}\selectfont  \@author\par}
\vskip25pt%
}%
}%
\makeatother

\begin{document}

\flushbottom
\maketitle
\thispagestyle{empty}

%\noindent Please note: Abbreviations should be introduced at the first mention in the main text – no abbreviations lists. Suggested structure of main text (not enforced) is provided below.

\section*{Abstract}

Highly magnetized neutron stars are promising candidates to explain some of the most peculiar astronomical phenomena, for instance, fast radio bursts, gamma-ray bursts, and superluminous supernovae \cite{2015ApJ...807..179P, 2019ApJ...886..110M, 2019ApJ...879....4W, 2020Natur.587...45Z, 2018MNRAS.481.2407M, 2008MNRAS.383L..25B, 2004ApJ...611..380T, 2012MNRAS.426L..76D}.
Pulsations of these highly magnetized neutron stars are also speculated to produce detectable gravitational waves. 
In addition, pulsations are important probes of the structure and equation of state of the neutron stars.
The major challenge in studying the pulsations of highly magnetized neutron stars is the demanding numerical cost of consistently solving the nonlinear Einstein and Maxwell equations under minimum assumptions. 
With the recent breakthroughs in numerical solvers \cite{XNS3, 2020CQGra..37n5015C}, we investigate pulsation modes of non-rotating neutron stars which harbour strong purely toroidal magnetic fields of $10^{15-17}$ G through two-dimensional axisymmetric general-relativistic magnetohydrodynamics simulations.
We show that stellar oscillations are insensitive to magnetization effects until the magnetic to binding energy ratio goes beyond 10\%, where the pulsation mode frequencies are strongly suppressed.
We further show that this is the direct consequence of the decrease in stellar compactness when the extreme magnetic fields introduce strong deformations of the neutron stars.

\section*{Introduction}

%The Introduction section, of referenced text\cite{Figueredo:2009dg} expands on the background of the work (some overlap with the Abstract is acceptable). The introduction should not include subheadings.

Neutron stars (NSs) are compact objects formed by core-collapse supernovae.
Due to field amplification in the violent formation processes, most NSs are endowed with strong magnetic fields of $10^{11-13}$ G \cite{1990CAS....16.....L}.
In some extreme cases, the magnetars can harbour even stronger magnetic fields of $10^{14-16}$ G, about 1000 times stronger than usual pulsars (for comparison, the magnetic field of a sunspot is $10^{3}$ G \cite{2013ARA&A..51..311S}).
Younger magnetars may carry even higher magnetic fields since they have been subjected to dissipative processes for shorter times \cite{2014MNRAS.439.3541P}.

These extreme magnetic fields affect the structure and evolution of NSs.
For instance, strong magnetic fields can deform NSs \cite{2008MNRAS.385..531H, 2014MNRAS.439.3541P}.
A direct consequence of structural deformations of NSs could be significant gravitational wave emissions \cite{1996A&A...312..675B, 2002PhRvD..66h4025C, 2011MNRAS.417.2288M}.
The geometry of magnetic fields of the NSs is a crucial factor governing the physics of NSs. However, the field configuration inside the NS is unknown.
Studies of equilibrium models with simple field configurations suggest that a purely toroidal field makes NSs prolate \cite{2008PhRvD..78d4045K,2009ApJ...698..541K,2012MNRAS.427.3406F} while a purely poloidal field forces the stars to become oblate \cite{1995A&A...301..757B,2001A&A...372..594K,2012PhRvD..85d4030Y}.
Nevertheless, these simple geometries are expected to be unstable \cite{1956ApJ...123..498P,1973MNRAS.161..365T,1973MNRAS.162..339W,1973MNRAS.163...77M,1974MNRAS.168..505M}.
Numerical simulations suggest that the magnetic fields of the NSs are rearranged rapidly due to these instabilities, leading to a mixed configuration of toroidal and poloidal fields, which is roughly axisymmetric \cite{2006A&A...450.1077B,2006A&A...450.1097B,2009MNRAS.397..763B,2012ApJ...760....1C}.
This mixed geometry is usually called \emph{twisted torus}.

Pulsations of NSs could be excited by various astrophysical events, such as core-collapse supernova and giant flares \cite{2018ASSL..457.....R}.
These pulsations are potential sources of gravitational waves, the spectra of which may serve as a sensitive probe of the structure and the equation of state (EoS) of NSs. 
Oscillation modes of non-magnetized NSs have been well studied using either perturbative calculations or dynamical simulations with or without spacetime evolutions, e.g. Refs. \cite{1998MNRAS.299.1059A,1999ApJ...515..414Y,2001MNRAS.320..307K,2001MNRAS.325.1463F,2002ApJ...568L..41Y,2002PhRvD..65h4024F,2004PhRvD..70l4015B,2005MNRAS.356..217Y,2010PhRvD..81h4019K,2011PhRvD..83f4031G,2013PhRvD..88d4052D}. Magnetic fields are also considered in studies based on either Newtonian approaches, e.g. Refs. {\cite{2010MNRAS.405..318L,10.1111/j.1365-2966.2010.17499.x,2011MNRAS.412.1730L,2015MNRAS.449.3620A,2016MNRAS.455.2228A}} or general-relativistic approaches with Cowling approximation (evolving matter equations only while keeping the spacetime fixed), e.g. Refs. {\cite{2001MNRAS.328.1161M,2007JPhCS..68a2049K,2009MNRAS.395.1163S,2010JPhCS.229a2079S,2011MNRAS.414.3014C,2012MNRAS.421.2054G,2012MNRAS.420.3035V,2013MNRAS.430.1811G}}. However, it has been shown that simulations using the Cowling approximation can overestimate the oscillation frequency up to a factor of 2 {\cite{10.1111/j.1365-2966.2004.07973.x,2006MNRAS.368.1609D}}. Therefore, it is important that, when computationally feasible, simulations with dynamical spacetime are conducted.

The major difficulties in studying magnetized NSs come from the non-linear nature of Einstein equations, and with Maxwell equations fully coupled, analytical calculations are generally impossible. Hence, numerical computations are inevitable to solve all the involved physics with a minimum number of assumptions. Until recently, due to breakthroughs in general-relativistic magnetohydrodynamics (GRMHD) simulations, dynamical studies of magnetized NSs have become possible, e.g. Refs. \cite{2011ApJ...735L..20L, 2011MNRAS.417.2288M, 2012PhRvD..85b4030Z, 2012ApJ...760....1C, 2013MNRAS.431.1853P, 2013PhRvD..88d4020D, 2013PhRvD..88j3005L, 2018PPCF...60a4027D}. Nonetheless, there is still no accurate eigenfrequency determination for oscillation modes in highly magnetized NSs.

A novel approach to compute strongly magnetized equilibrium models is recently presented and demonstrated by an open-source code \texttt{XNS} \cite{2011A&A...528A.101B,2014MNRAS.439.3541P,XNS2,XNS3,2014IJMPS..2860202P,2015MNRAS.447.3278B,2016JPhCS.719a2004B,2016MNRAS.462L..26P,2017MNRAS.470.2469P,2020A&A...640A..44S,2021A&A...654A.162S,2021A&A...645A..39S}. Moreover, the GRMHD code \texttt{Gmunu} \cite{2020CQGra..37n5015C,2021ApJ...915..108N,2021MNRAS.508.2279C,2022ApJS..261...22C} allows us to robustly evolve NSs in dynamical spacetime even with extreme magnetic fields of $10^{15-17}$ G. With these powerful tools in hand, we are now in a much better position to systematically investigate the oscillation modes of magnetized NSs.

In this work, we numerically study the oscillations of highly magnetized non-rotating axisymmetric (two-dimensional) NSs.
Specifically, we first construct 12 equilibrium models with different magnetic to binding energy ratios $\mathscr{H}/\mathscr{W}$ using \texttt{XNS}, including one non-magnetized reference model named `REF' and 11 magnetized models in ascending order of $\mathscr{H}/\mathscr{W}$ named T1K1, T1K2, ..., T1K11 (Methods).
Next, we utilize \texttt{Gmunu} to perturb and evolve the equilibrium models in dynamical spacetime, where we try three different initial fluid perturbations for excitation of stellar oscillations, namely $\ell=0$, $\ell=2$, and $\ell=4$ perturbations (Methods).
After that, we perform a Fourier analysis of the simulation results to examine how the eigenfrequencies of oscillation modes vary with $\mathscr{H}/\mathscr{W}$ of the NS (Methods), and we discuss possible reasons behind our results.

%Up to three levels of \textbf{subheading} are permitted. Subheadings should not be numbered.

%\subsection*{Subsection}

%Example text under a subsection. Bulleted lists may be used where appropriate, e.g.

%\begin{itemize}
%\item First item
%\item Second item
%\end{itemize}

%\subsubsection*{Third-level section}
 
%Topical subheadings are allowed.

\section*{Results}
\subsection*{\label{sec:mag_vs_f}Magnetization effects on oscillations of NSs}
\begin{figure}[ht]
	\centering
	\includegraphics[width=\linewidth]{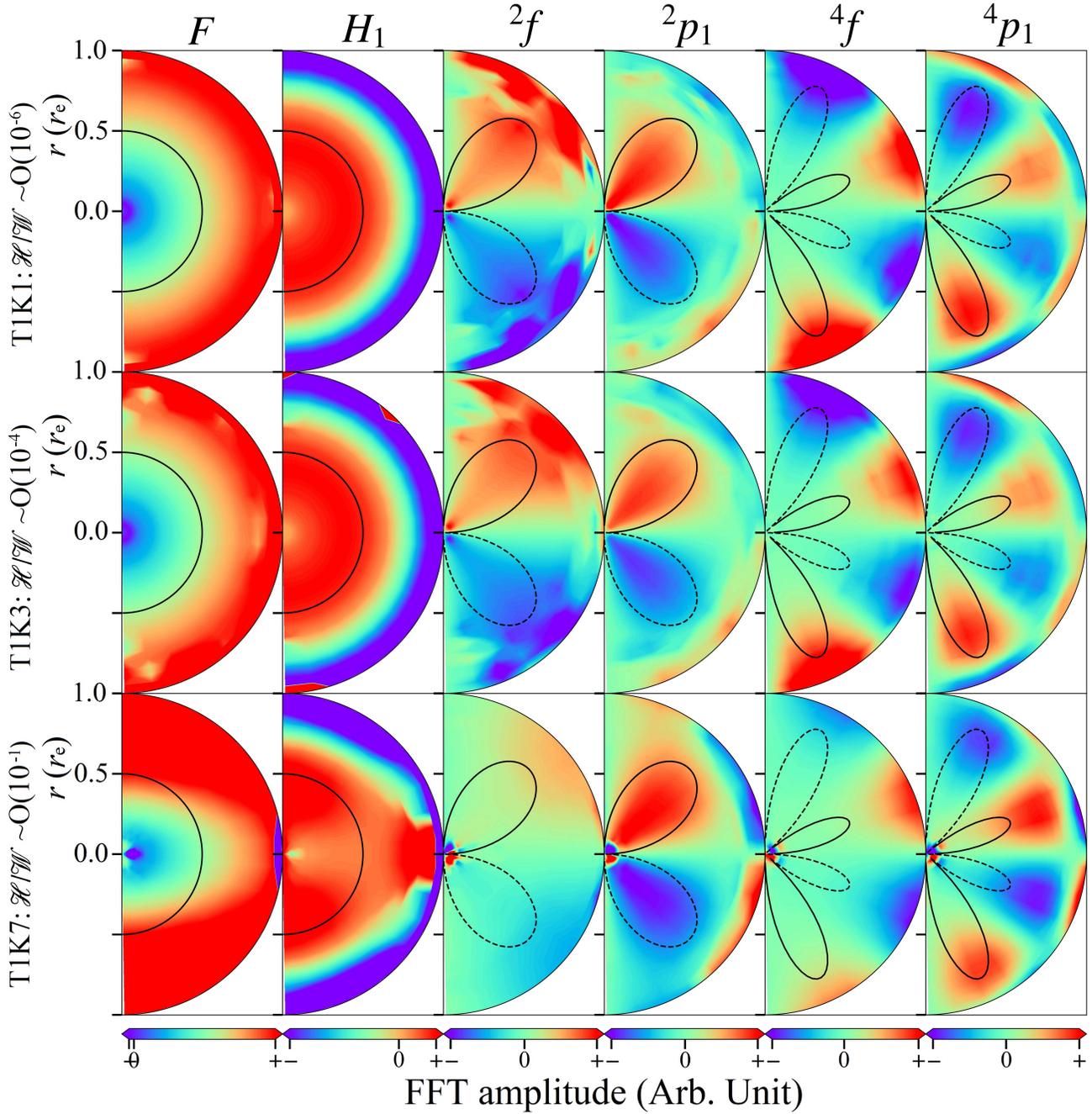}
	\caption{\label{fig1} Visualizations of eigenfunctions of the six dominant oscillation modes using the data of 3 equilibrium models (T1K1, T1K3 and T1K7).
	The fundamental quasi-radial $(\ell=0)$ mode $F$ and its first overtone $H_1$ are predominantly excited under $\ell=0$ perturbation; the fundamental quadrupole $(\ell=2)$ mode $^2f$ and its first overtone $^2p_1$ are predominantly excited under $\ell=2$ perturbation; the fundamental hexadecapole $(\ell=4)$ mode $^4f$ and its first overtone $^4p_1$ are predominantly excited under $\ell=4$ perturbation.
	Each polar color plot shows the spatial map of FFT amplitude at the eigenfrequency of the mode, where the radial axis is normalized to the equatorial radius $r_{\textrm e}$ of each model.
	On top of each color plot, there is a polar line plot visualizing the $\theta$-part of the spherical harmonic in the corresponding perturbation function, where the distance from the origin to the line measures the magnitude of the spherical harmonic in that $\theta$-direction, while the solid and dotted portions represent the positive and negative parts of the spherical harmonic respectively.
	Each line plot is scaled arbitrarily for clearer illustration.
	It can be seen that the eigenfunctions of the higher-order quadrupole $(\ell=2)$ and hexadecapole $(\ell=4)$ modes have more nodes in the $\theta$-direction compared to the quasi-radial $(\ell=0)$ modes, while the eigenfunction of each first overtone has more nodes in the $r$-direction compared to its fundamental mode.
	Furthermore, each eigenfunction qualitatively agrees with the spherical harmonic in the corresponding perturbation function.}
\end{figure}

\begin{figure}[ht]
	\centering
	\includegraphics[height=0.8\textheight]{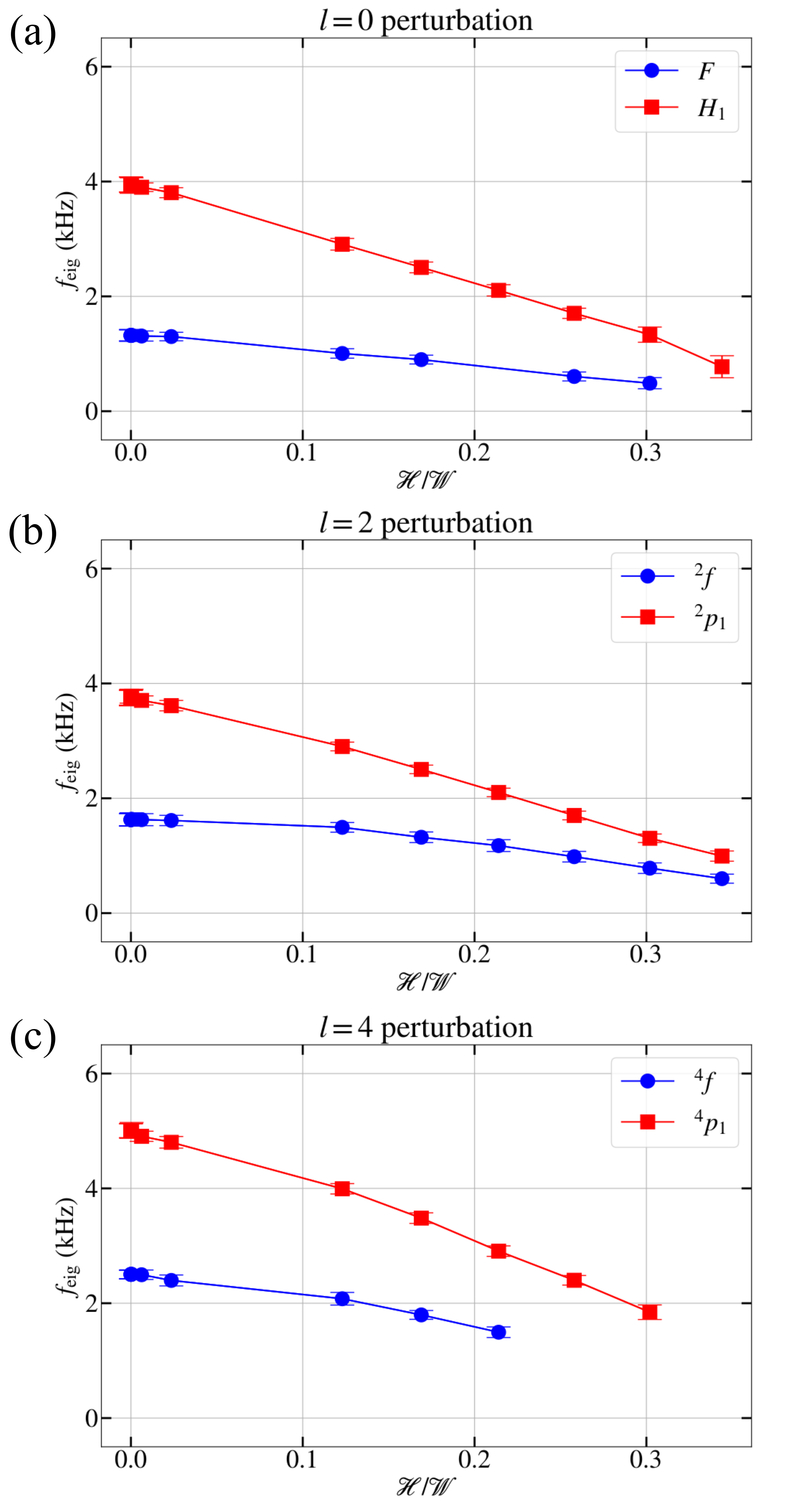}
	\caption{\label{fig2} Plots of eigenfrequencies $f_{\textrm{eig}}$ of the excited oscillation modes against the magnetic to binding energy ratio $\mathscr{H}/\mathscr{W}$ of the NS model, if $\ell=0$ (\emph{Upper panel}), $\ell=2$ (\emph{Middle panel}), and $\ell=4$ (\emph{Lower panel}) perturbations are applied respectively.
	For all the modes, the data points at $\mathscr{H}/\mathscr{W} \sim 0$ (corresponding to REF - T1K5 models) show a nearly horizontal trend, even though these models span a few orders of magnitude in $\mathscr{H}/\mathscr{W}$ and can achieve a maximum field strength of $10^{15-17}$ G.
	This implies magnetization has negligible effects on stellar oscillations for NSs with $\mathscr{H}/\mathscr{W} \lesssim 10^{-2}$.
	However, $f_{\textrm{eig}}$ noticeably decreases with $\mathscr{H}/\mathscr{W}$ for $\mathscr{H}/\mathscr{W} \gtrsim 10^{-1}$ in general, and as explained in the caption of Table~\ref{table1}, the expected higher-order hexadecapole $(\ell=4)$ modes are suppressed or even disappear in the most magnetized models under $\ell=4$ perturbation.
	Hence, we can see that stellar oscillations are significantly suppressed by stronger magnetization for NSs with $\mathscr{H}/\mathscr{W} \gtrsim 10^{-1}$.
	}
\end{figure}

In total, six dominant oscillation modes are observed in our numerical study, namely the fundamental quasi-radial $(\ell=0)$ mode $F$ and its first overtone $H_1$, the fundamental quadrupole $(\ell=2)$ mode $^2f$ and its first overtone $^2p_1$, as well as the fundamental hexadecapole $(\ell=4)$ mode $^4f$ and its first overtone $^4p_1$ (we follow the notations in Ref.~\cite{2006MNRAS.368.1609D}).
Each mode is predominantly excited under the initial perturbation with the corresponding $\ell$ index, and each eigenfunction qualitatively agrees with the spherical harmonic in the corresponding perturbation function, as shown in Fig.~\ref{fig1}.
The measured eigenfrequencies of the six modes in the 12 different NS models are summarized in Table~\ref{table1}, where the undetermined eigenfrequencies denoted by `N/A' in different columns stem from different reasons below.
For the column of $F$ mode, the missing eigenfrequencies are due to unsatisfactory data quality in \texttt{Gmunu} simulations of T1K8 and T1K11 models under $\ell=0$ perturbation.
On the other hand, for the columns of $^4f$ and $^4p_1$ modes, some eigenfrequencies are missing because the hexadecapole $(\ell=4)$ modes are masked by the quadrupole $(\ell=2)$ modes and are no longer the dominant modes in \texttt{Gmunu} simulations of the most magnetized models under $\ell=4$ perturbation.
To better illustrate the results in Table~\ref{table1}, we plot in Fig.~\ref{fig2} the eigenfrequencies $f_{\textrm{eig}}$ of the six modes as functions of the magnetic to binding energy ratio $\mathscr{H}/\mathscr{W}$ of the NS model.

We have observed an $\mathscr{H}/\mathscr{W}$ threshold for stellar magnetization to start affecting the oscillations of NSs.
For NSs with $\mathscr{H}/\mathscr{W} \lesssim 10^{-2}$, stellar oscillations are insensitive to magnetization effects.
This can be seen from Table~\ref{table1} that $f_{\textrm{eig}}$ of every oscillation mode is nearly the same for the first six models (REF - T1K5) even though these models span a few orders of magnitude in $\mathscr{H}/\mathscr{W}$ and can achieve a maximum field strength of $10^{15-17}$ G; this can also be seen from Fig.~\ref{fig2} that the data points at $\mathscr{H}/\mathscr{W} \sim 0$ show a nearly horizontal trend.
On the other hand, for NSs with $\mathscr{H}/\mathscr{W} \gtrsim 10^{-1}$, stellar oscillations are significantly suppressed by stronger magnetization.
Refer to the data points at $\mathscr{H}/\mathscr{W} > 10^{-1}$ in Fig.~\ref{fig2}, $f_{\textrm{eig}}$ decreases with $\mathscr{H}/\mathscr{W}$ in general, and all the oscillation modes are pushed towards the low-frequency region, leading to the near-degeneracy of $H_1$ and $^2p_1$ modes.
Moreover, as afore-explained about the undetermined eigenfrequencies, $\ell=4$ perturbation excites the quadrupole $(\ell=2)$ modes preferentially over the expected hexadecapole $(\ell=4)$ modes in the most magnetized models, hinting at suppression or even disappearance of higher-order oscillation modes in a more magnetized NS for $\mathscr{H}/\mathscr{W} \gtrsim 10^{-1}$.
To summarize, magnetization effects start to hinder stellar oscillations if $\mathscr{H}/\mathscr{W}$ of the NS passes the threshold somewhere between $10^{-2}$ and $10^{-1}$.

\subsection*{\label{sec:comp_vs_f}Compactness as an underlying factor}
\begin{figure}[ht]
	\centering
	\includegraphics[width=\linewidth]{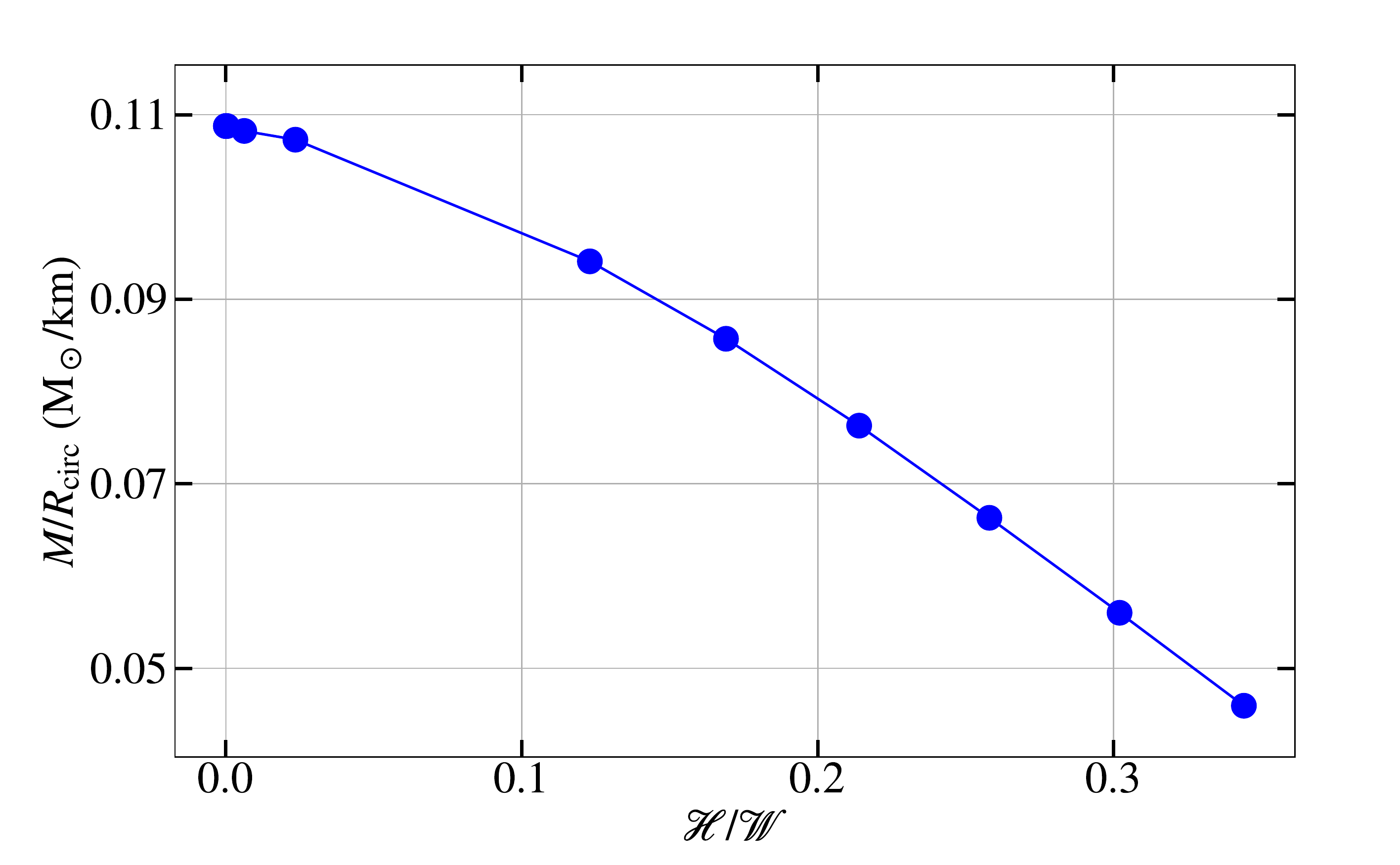}
	\caption{\label{fig3} Plot of compactness $M/R_{\textrm circ}$ against the magnetic to binding energy ratio $\mathscr{H}/\mathscr{W}$ of the NS model, where $M$ is the gravitational mass and $R_{\textrm circ}$ is the circumferential radius.
	$M/R_{\rm circ}$ remains nearly unchanged for $\mathscr{H}/\mathscr{W} \lesssim 10^{-2}$ but decreases dramatically for $\mathscr{H}/\mathscr{W} > 10^{-1}$.
	This agrees with the trends of $f_{\textrm{eig}}(\mathscr{H}/\mathscr{W})$ shown in Fig.~\ref{fig2} and reveals a correlation between eigenfrequencies of oscillation modes and stellar compactness for magnetized NSs.}
\end{figure}

\begin{figure}[ht]
	\centering
	\includegraphics[width=\linewidth]{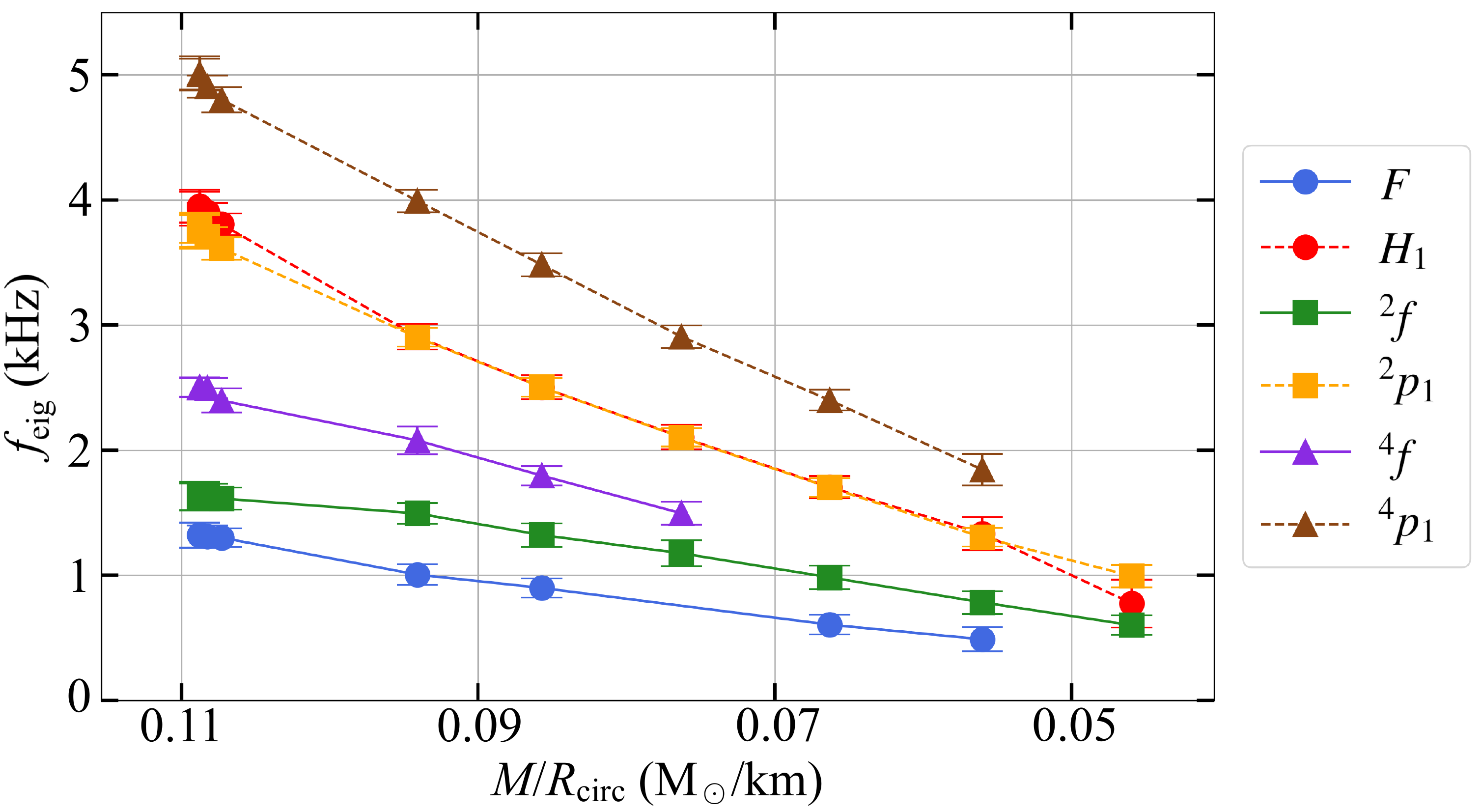}
	\caption{\label{fig4} Plot of eigenfrequencies $f_{\textrm{eig}}$ of the excited oscillation modes against compactness $M/R_{\textrm circ}$ of the NS model, where $M$ is the gravitational mass and $R_{\textrm circ}$ is the circumferential radius.
	All data points show that $f_{\textrm{eig}}$ decreases together with $M/R_{\rm circ}$ in an almost linear way. Hence, this demonstrates a quasilinear relation between eigenfrequencies of oscillation modes and stellar compactness for magnetized NSs..
	}
\end{figure}

The magnetization effects on NS oscillations discussed above may be understood by studying the compactness $M/R_{\rm circ}$ of the NS, where $M$ is the gravitational mass and $R_{\rm{circ}}$ is the circumferential radius.
As shown Ref. \cite{1975ApJ...196..653H}, the eigenfrequencies of the fundamental quasi-radial and quadrupole modes are related to the stellar compactness for non-magnetized NSs, and we suspect this correlation also holds for highly magnetized NSs.
Thus, based on our NS models, we plot in Fig.~\ref{fig3} the compactness $M/R_{\rm circ}$ against the magnetic to binding energy ratio $\mathscr{H}/\mathscr{W}$.
We find that $M/R_{\rm circ}$ remains nearly unchanged for $\mathscr{H}/\mathscr{W} \lesssim 10^{-2}$ but decreases dramatically for $\mathscr{H}/\mathscr{W} > 10^{-1}$, which agrees with the trends of $f_{\textrm{eig}}(\mathscr{H}/\mathscr{W})$ shown in Fig.~\ref{fig2} and indeed reveals a correlation between eigenfrequencies of oscillation modes and stellar compactness.
We also plot in Fig.~{\ref{fig4}} $f_{\textrm{eig}}$ against $M/R_{\rm circ}$. For all the modes, $f_{\textrm{eig}}$ decreases together with $M/R_{\rm circ}$ in an almost linear way. Therefore, we found a quasilinear relation between $f_{\textrm{eig}}$ and $M/R_{\rm circ}$ for magnetized NSs. The complete physical interpretation of our results is that a strong toroidal field can cause deformation of the NS \cite{2014MNRAS.439.3541P} and alter the stellar compactness, so the propagation of seismic activities inside the NS is affected. In consequence, the eigenfrequencies of oscillation modes are correspondingly modified.

\section*{Discussion}

%The Discussion should be succinct and must not contain subheadings.

In this work, we systematically investigate how a strong purely toroidal magnetic field with a field strength of $10^{15-17}$ G affects the oscillations of non-rotating NSs via two-dimensional axisymmetric simulations.
We carefully extract the eigenfrequencies of the excited oscillation modes and construct the corresponding eigenfunctions from the simulated data.
We have found that stellar oscillations are insensitive to magnetization effects for NSs with magnetic to binding energy ratio $\mathscr{H}/\mathscr{W} \lesssim 10^{-2}$, even though the maximum magnetic field strength $B_{\max}$ can reach $\mathcal{O}(10^{17})$ G in the star.
However, stellar oscillations are suppressed significantly by stronger magnetization if $\mathscr{H}/\mathscr{W} \gtrsim 10^{-1}$.
This behaviour can be understood by the decrease of stellar compactness due to strong magnetic fields.
We show that the compactness has the same dependence on $\mathscr{H}/\mathscr{W}$ as the eigenfrequencies and demonstrate that the correlation between eigenfrequencies and compactness exists not only in non-magnetized NSs \cite{1975ApJ...196..653H} but also in highly magnetized NSs.

We compare our results with previous Newtonian studies, e.g. Ref. {\cite{2010MNRAS.405..318L,2015MNRAS.449.3620A}}. These studies considered either perturbative or self-consistent MHD to construct the equilibrium models in the Newtonian regime. Both approaches found that the magnetic distortion and frequency shift in oscillation modes due to toroidal fields are minor corrections approximately proportional to $B^2$ (or roughly $\mathscr{H}/\mathscr{W}$ in this work).
However, in our GRMHD simulations, the equilibrium models are constructed by solving self-consistent general-relativistic magnetohydrostatic equations in the code \texttt{XNS} {\cite{2011A&A...528A.101B,2014MNRAS.439.3541P,XNS2,XNS3,2014IJMPS..2860202P,2015MNRAS.447.3278B,2016JPhCS.719a2004B,2016MNRAS.462L..26P,2017MNRAS.470.2469P,2020A&A...640A..44S,2021A&A...654A.162S,2021A&A...645A..39S}}. When $\mathscr{H}/\mathscr{W} \gtrsim 10^{-1}$, the magnetic deformations are far from small corrections, and thus the stellar compactness is significantly reduced. Therefore, the effect of decreasing compactness dominates and results in the suppression of oscillation modes.  
Besides, we compare our results with those under the Cowling approximation and we corroborate what has been shown in the literature {\cite{10.1111/j.1365-2966.2004.07973.x,2006MNRAS.368.1609D}}, namely that the Cowling approximation can lead to errors of factors of 2 (see Supplementary information).

The strongest magnetic field strength of $10^{17}$ G in this work is not expected to be observed in the exterior of ordinary pulsars and magnetars. Nevertheless, since the toroidal fields are enclosed inside the NSs, this ultra-high field could exist in the interior regions. Moreover, such field strength could also be generated during the formation of a proto-NS {\cite{2014MNRAS.439.3541P}}, and binary neutron star mergers {\cite{2006Sci...312..719P}}. The excited oscillation modes in these scenarios are potential sources for gravitational waves, and they could be detected with the next-generation detectors, such as the Kamioka Gravitational Wave Detector (KAGRA) {\cite{2018LRR....21....3A}}, the Einstein Telescope (ET) {\cite{2010CQGra..27s4002P}}, and the Neutron Star Extreme Matter Observatory (NEMO) {\cite{2020PASA...37...47A}}.

This work presents the first step to understanding how magnetic fields with different geometries affect the oscillations of NSs.
Since stellar models with purely toroidal fields are generally unstable {\cite{1973MNRAS.161..365T}}, the instability is only suppressed due to the restriction to 2D axisymmetry in this work. Therefore, a natural extension considers strong purely poloidal fields and the more realistic twisted torus configuration.
Since these field configurations extend to the regions outside NSs, an accurate and robust resistive GRMHD solver could be used to model these regions. This solver has already been implemented into \texttt{Gmunu} \cite{2022ApJS..261...22C} for future studies.
In addition to different configurations of magnetic fields, rotation should also be taken into account to work towards a more realistic problem, as the observed NSs are suggested to be rotating.
Furthermore, introducing realistic EoSs is essential since one of the most important purposes of oscillation studies is to probe the structure and the EoSs of NSs.

\section*{Methods}
%Topical subheadings are allowed. Authors must ensure that their Methods section includes adequate experimental and characterization data necessary for others in the field to reproduce their work.

\subsection*{\label{sec:models}Equilibrium models}
Equilibrium models of NSs are constructed by the code \texttt{XNS} {\cite{2011A&A...528A.101B,2014MNRAS.439.3541P,XNS2,XNS3,2014IJMPS..2860202P,2015MNRAS.447.3278B,2016JPhCS.719a2004B,2016MNRAS.462L..26P,2017MNRAS.470.2469P,2020A&A...640A..44S,2021A&A...654A.162S,2021A&A...645A..39S}}. \texttt{XNS} is a branch of the X-ECHO code {\cite{2011A&A...528A.101B}} developed to compute equilibrium models of highly magnetized axisymmetric NSs with rotations. Different magnetic field configurations {\cite{2014MNRAS.439.3541P}}, uniformly and differentially rotating profiles {\cite{2011A&A...528A.101B}}, and polytropic and non-polytropic tabulated equations of state {\cite{2021A&A...654A.162S}} are admitted. \texttt{XNS} enforces the $3+1$ formulism, the conformal flatness condition, and the assumption of axisymmetric and stationary space-time so that the line element can be written as
\begin{linenomath*}
    \begin{equation} \label{ds^2}
	    ds^2 = - \alpha ^2 dt^2 + \psi^4 \left[ dr^2 + r^2 d\theta^2 + r^2\sin^2\theta \left( d\phi + \beta^\phi dt\right)^2\right],
    \end{equation}
\end{linenomath*}
where $\alpha(r,\theta)$ is the lapse function, $\psi(r,\theta)$ is the conformal factor, and $\beta^{\phi}(r,\theta)$ is the shift vector ($\beta^{\phi}=0$ for non-rotating configurations). 

We assume a polytropic EoS $p=K\rho^{\gamma}$ for the stellar fluid, where $p$ is the pressure, $K$ is the polytropic constant, $\rho$ is the density, and $\gamma$ is the adiabatic index; as well as a polytropic expression $B_{\phi}=\alpha^{-1}K_{\textrm m}(\rho h\varpi^2)^m$ for the toroidal field, where $K_{\textrm m}$ is the toroidal magnetization constant, $h$ is the specific enthalpy, $\varpi^2=\alpha^2\psi^4r^2\sin^2\theta$, and $m\geq1$ is the toroidal magnetization index. Although the field configuration of an isolated NS is expected to be a mixture of toroidal and poloidal fields, it is important first to assess how a simpler field geometry would affect the oscillations of NSs before we move on to the more complicated \emph{Twisted Torus} case.

In total, 12 equilibrium models are computed with \texttt{XNS}, where one of them is a non-magnetized reference model named `REF', and the remaining 11 models are magnetized.
All the 12 models share the same rest mass $M_0=1.68$ M$_\odot$, and the same $K=1.6 \times 10^5$ cm$^5$ g$^{-1}$ s$^{-2}$ and $\gamma=2$ in the fluid EoS.
The 11 magnetized models have the same $m=1$ but different values of $K_{\textrm m}$ in the $B_{\phi}$ expression, and they are arranged in ascending order of magnetic to binding energy ratio $\mathscr{H}/\mathscr{W}$, where the one with the lowest ratio is named `T1K1', and the one with the second-lowest ratio is named `T1K2', so on and so forth.
(`T1' specifies the toroidal magnetization index being 1 and `K' stands for $K_{\textrm m}$).
The detailed properties of all the 12 models are summarized in Table~\ref{table2}.

\subsection*{\label{sec:perturbations}Initial perturbations to excite oscillations}
Consulting a similar study on rotating non-magnetized NSs done by Ref. \cite{2006MNRAS.368.1609D}, we try the following three types of initial fluid perturbations for exciting oscillations in the equilibrium models. 

First, we have the $\ell=0$ perturbation on the $r$-component of the three-velocity field,
\begin{linenomath*}
    \begin{equation}
        \delta v^r=a \sin\left[\pi\frac{r}{r_{\textrm{s}}(\theta)}\right],
    \label{l0} 
    \end{equation}
\end{linenomath*}
where $r_{\textrm{s}}(\theta)$ locates the surface of the NS, and the perturbation amplitude $a$ (in unit of c) is chosen to be 0.001.

Second, we have the $\ell=2$ perturbation on the $\theta$-component of the three-velocity field,
\begin{linenomath*}
    \begin{equation}
        \delta v^{\theta}=a \sin\left[\pi\frac{r}{r_{\textrm{s}}(\theta)}\right] \sin\theta \cos\theta,
    \label{l2} 
    \end{equation}
\end{linenomath*}
where $a$ is chosen to be 0.01.

Lastly, we have the $\ell=4$ perturbation on the $\theta$-component of the three-velocity field,
\begin{linenomath*}
    \begin{equation}
        \delta v^{\theta}=a \sin\left[\pi\frac{r}{r_{\textrm{s}}(\theta)}\right] \sin\theta \cos\theta (3-7\cos^2\theta),
    \label{l4} 
    \end{equation}
\end{linenomath*}
where $a$ is again set to be 0.01.

All the three perturbation functions comprise a sine function of $r$ and the $\theta$-part of a spherical harmonic with the corresponding $\ell$ index.
The sine function of $r$ has its nodes at the centre and on the surface of the NS to avoid initial perturbations on sensitive boundaries of the problem and minimize any potential numerical errors.
On the other hand, spherical harmonics are a natural choice for exciting oscillations on a sphere-like object.
Moreover, for the higher-order $\ell=2$ and $\ell=4$ perturbations, the perturbation amplitude $a$ has to be larger to induce any observable oscillations.

\subsection*{\label{sec:simulations}Simulations}
Simulations are performed with our code \texttt{Gmunu} \cite{2020CQGra..37n5015C,2021ApJ...915..108N,2021MNRAS.508.2279C,2022ApJS..261...22C}. 
For each of the 12 equilibrium models, we execute \texttt{Gmunu} three times, once for each initial perturbation function.
Hence, $12 \times 3 = 36$ simulations are carried out in total.
In all the 36 simulations, the models are evolved over a time span of 10 ms with the polytropic EoS $p = K \rho^{\gamma}$, under the same setting as in the computation of equilibrium models, namely, $\gamma = 2$ and $K=110$.
The lowest allowed rest mass density (`atmosphere') is set to be $\rho_\text{atmo} = \rho_{\max}\left( t=0\right) \times 10^{-10}$, and the ratio of $\rho_\text{atmo}$ to threshold density $\rho_\text{thr}$ is $\rho_\text{atmo} / \rho_\text{thr} = 0.99$.
For completeness, we also perform simulations under the Cowling approximation (see Supplementary information), with other settings unchanged.

The two-dimensional computational domain covers $0 \leq r \leq 60$, $0 \leq \theta \leq \pi$ with the resolution $N_r \times N_\theta = 64 \times 16$ where each block has $8^2$ cells, thus allowing 4 AMR level (an effective resolution of $512 \times 128$).
The grid refinement used in this study is identical to the GR simulations in Ref. \cite{2021MNRAS.508.2279C}.
In particular, we define a relativistic gravitational potential $\Phi := 1 - \alpha$. As $\Phi$ is almost proportional to $M/R$, we can use $\Phi^{-1}$ as a measure of the characteristic length-scale {\cite{2021MNRAS.508.2279C}}.
For any $\Phi$ larger than the maximum potential $\Phi_{\text{max}}$ (which is set as 0.2 in this work), the block is set to be the finest.
While for the second-finest level, the same check is performed with a new maximum potential which is half of the previous one, so on and so forth.
To avoid the rigorous Courant-Friedrichs-Lewy (CFL) condition at the centre of the star, the grids are enforced to be coarsened for keeping $r \Delta \theta \sim \Delta r$ when $r$ is smaller than 0.5. 
(Unless otherwise specified, all quantities in this subsection are in dimensionless units $c = G = M_{\odot} = 1$.)

\subsection*{\label{sec:extract_f}Extraction of eigenfrequencies and eigenfunctions}
We analyze the data from a \texttt{Gmunu} simulation in the following three steps.
For the first step, we extract the time evolutions of the initially perturbed component of the three-velocity field at 361 $(r, \theta)$-points in the NS model and compute the Fast Fourier Transform (FFT) of the temporal data at each $(r, \theta)$-point.
Hence, 361 FFT spectra, plots of magnitude of the complex FFT in the frequency domain, are obtained altogether.
According to Ref.~\cite{2006MNRAS.368.1609D}, our initial perturbation amplitudes are small enough such that the overall evolution of the input model in a \texttt{Gmunu} simulation can be described as a superposition of a few global oscillation modes.
We verify this by observing that the FFT spectra obtained at different spatial points show discrete peaks and agree well on the peak positions.

\begin{figure}[ht]
	\centering
	\includegraphics[width=\linewidth]{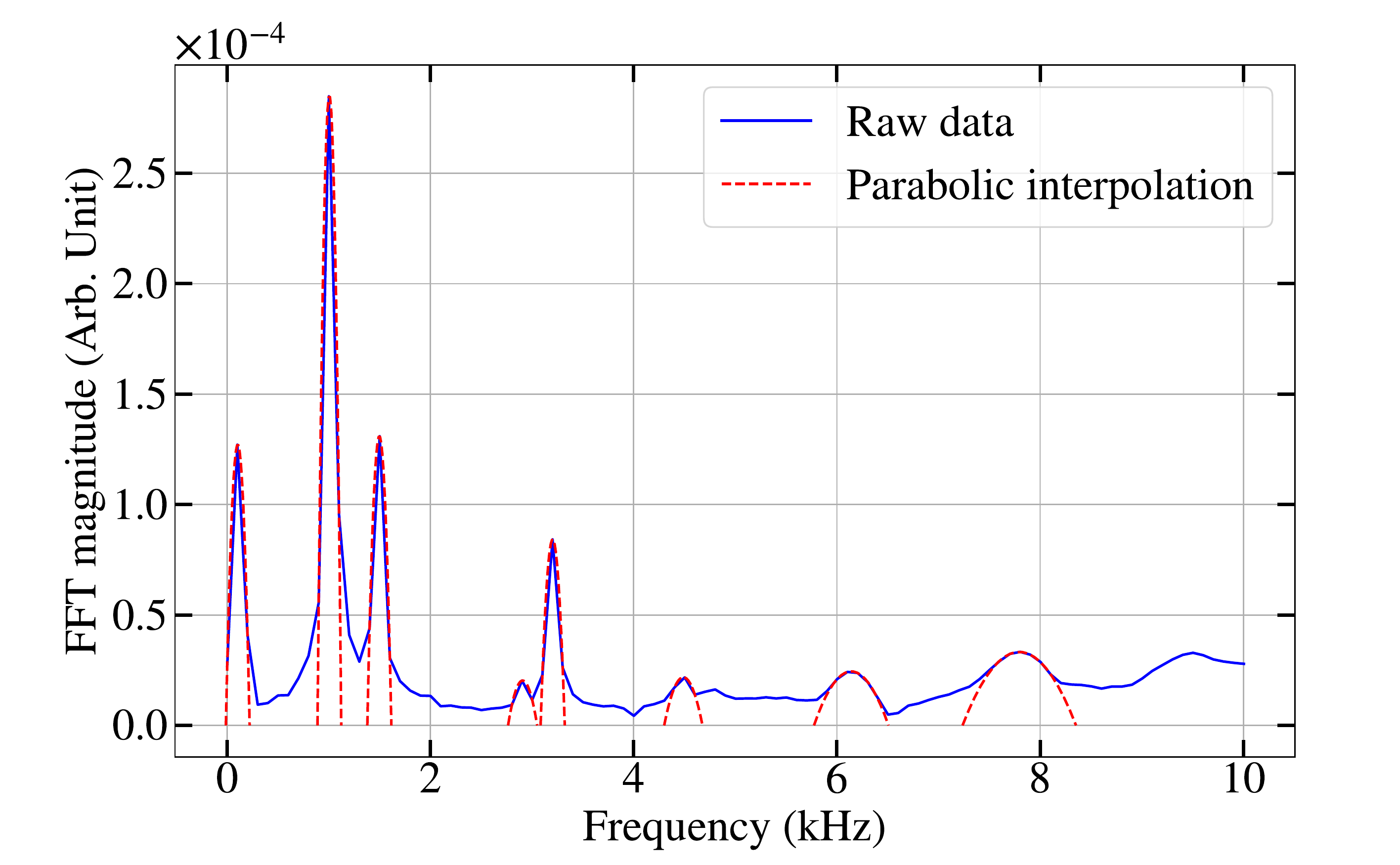}
	\caption{\label{fig5} Single-point FFT spectrum obtained from the \texttt{Gmunu} simulation of T1K6 model under $\ell=0$ perturbation.
	The FFT data are computed from the time evolution of the $r$-component of three-velocity field at the spatial point $(r, \theta) \simeq (r_{\textrm e}/2, \pi/2)$.
	The parabolic interpolations of the prominent peaks are shown as an example.
	The FFT spectrum at a single spatial point usually shows peaks that are sharp enough for analysis.}
\end{figure}

For the second step, we extract the eigenfrequencies of the excited oscillation modes.
Usually, the FFT spectrum at a spatial point where the initial perturbation function has a large magnitude can reveal FFT peaks loud enough for further analysis (e.g. at $(r, \theta) \simeq (r_{\textrm e}/2, \pi/2)$, $(r_{\textrm e}/2, \pi/4)$, and $(r_{\textrm e}/2, 2\pi/15)$ for $\ell=0$, $\ell=2$, $\ell=4$ perturbations respectively).
Nevertheless, occasionally, we may have to integrate the FFT spectra along a radial line for sharper FFT peaks (along $\theta=\pi/2$, $\pi/4$, and $2\pi/15$ for $\ell=0$, $\ell=2$, $\ell=4$ perturbations respectively).
Since our study here is in the ideal GRMHD regime with no physical damping of the oscillations, we apply parabolic interpolation instead of Lorentzian fitting to the peaks in the single-point or integrated FFT spectrum for simplicity (see Fig.~\ref{fig5} as an example).
We then take the interpolated peak positions as the measured eigenfrequencies $f_{\textrm{eig}}$ and the full-width-at-half-maximums (FWHMs) of the parabolic interpolations as the uncertainties in eigenfrequency extraction.

For the third step, we extract the eigenfunctions of the excited oscillation modes.
According to Refs.~\cite{2006MNRAS.368.1609D,2004MNRAS.352.1089S}, the eigenfunction of a mode is correlated to the spatial map of FFT amplitude at the eigenfrequency of the mode, where FFT amplitude is the magnitude of the FFT multiplied by the sign of its real part.
Using our FFT data computed at the 361 points, we spatially map the FFT amplitude at the frequency to which the measured eigenfrequency is the closest in the discretized frequency domain of our FFT analysis for simplicity.
The eigenfunction visualized by such a spatial map can serve as a unique trademark to help us identify the same oscillation mode excited in different \texttt{Gmunu} simulations so that we can investigate the dependence of eigenfrequency $f_{\textrm{eig}}$ of a particular mode on the magnetic to binding energy ratio $\mathscr{H}/\mathscr{W}$ of the input model.

In the end, we can obtain the curves of $f_{\textrm{eig}}(\mathscr{H}/\mathscr{W})$ for different oscillation modes to examine the magnetization effects on oscillations of NSs.
Lastly, we determine the correspondence between the modes found in our study and the modes in the literature by comparing the eigenfrequencies at zero magnetic energy, $f_{\textrm{eig}}(\mathscr{H}/\mathscr{W}=0)$, of the modes we found here with the mode frequencies previously reported for a non-magnetized non-rotating NS model with a similar gravitational mass \cite{2006MNRAS.368.1609D}.

\bibliography{references}

%\noindent LaTeX formats citations and references automatically using the bibliography records in your .bib file, which you can edit via the project menu. Use the cite command for an inline citation, e.g.  \cite{Hao:gidmaps:2014}.

%For data citations of datasets uploaded to e.g. \emph{figshare}, please use the \verb|howpublished| option in the bib entry to specify the platform and the link, as in the \verb|Hao:gidmaps:2014| example in the sample bibliography file.

\section*{Acknowledgements}

%Acknowledgements should be brief, and should not include thanks to anonymous referees and editors, or effusive comments. Grant or contribution numbers may be acknowledged.

We acknowledge the support of the CUHK Central High Performance Computing Cluster, on which the scaling tests in this work have been performed.
This work was partially supported by grants from the Research Grants Council of Hong Kong (Project No. CUHK 14306419), the Croucher Innovation Award from the Croucher Fundation Hong Kong, and the Direct Grant for Research from the Research Committee of The Chinese University of Hong Kong.

\section*{Author contributions statement}
%Must include all authors, identified by initials, for example:
%A.A. conceived the experiment(s),  A.A. and B.A. conducted the experiment(s), C.A. and D.A. analysed the results.  
M.Y.L. contributed to project planning, simulations, simulation analysis, interpretation of results, and manuscript preparation.
A.K.L.Y. contributed to the simulations, simulation analysis, interpretation of results, and manuscript preparation.
P.C.K.C. conceived the idea for the project and contributed to project planning and leadership, simulation code development, simulations, interpretation, and manuscript preparation.
T.G.F.L. led the group.
All authors reviewed the manuscript.

\section*{Data availability}
The data generated and analysed during this study are available from the corresponding author on reasonable request.

\section*{Competing interests statement}
The authors declare no competing interests.

%To include, in this order: \textbf{Accession codes} (where applicable); \textbf{Competing interests} (mandatory statement). 
%The corresponding author is responsible for submitting a \href{http://www.nature.com/srep/policies/index.html#competing}{competing interests statement} on behalf of all authors of the paper. This statement must be included in the submitted article file.

%\begin{figure}[ht]
%\centering
%\includegraphics[width=\linewidth]{stream}
%\caption{Legend (350 words max). Example legend text.}
%\label{fig:stream}
%\end{figure}
%
%\begin{table}[ht]
%\centering
%\begin{tabular}{|l|l|l|}
%\hline
%Condition & n & p \\
%\hline
%A & 5 & 0.1 \\
%\hline
%B & 10 & 0.01 \\
%\hline
%\end{tabular}
%\caption{\label{tab:example}Legend (350 words max). Example legend text.}
%\end{table}
%
%Figures and tables can be referenced in LaTeX using the ref command, e.g. Figure \ref{fig:stream} and Table \ref{tab:example}.

\begin{table}[ht]
	\centering
	\begin{tabular}{cccccccc}
		 Model & $F$ & $H_1$ & $^2f$ & $^2p_1$ & $^4f$ & $^4p_1$\\
		\hline
		 REF	&	1.32	&	3.95	&	1.63	&	3.75	&	2.50	&   5.00\\
		 T1K1	&	1.32	&	3.95	&	1.63	&	3.75	&	2.50    &   5.00\\
		 T1K2	&	1.32	&	3.95	&	1.63	&	3.75	&	2.50    &   5.01\\
		 T1K3	&	1.32	&	3.93	&	1.63	&	3.78	&	2.51	&	5.01\\
		 T1K4	&	1.31	&	3.90	&	1.63	&	3.70	&	2.50	&	4.91\\
		 T1K5	&	1.30	&	3.81	&	1.61	&	3.61	&	2.40	&	4.80\\
		 T1K6	&	1.01	&	2.91	&	1.49	&	2.90	&	2.08	&	3.99\\
		 T1K7	&	0.90	&	2.50	&	1.32	&	2.50	&	1.80	&	3.48\\
		 T1K8	&	N/A     &	2.10	&	1.18	&	2.10	&	1.49	&	2.91\\
		 T1K9	&	0.60	&	1.70	&	0.98	&	1.70	&	N/A     &	2.40\\
		 T1K10	&	0.49	&	1.33	&	0.78	&	1.30	&	N/A	    &	1.84\\
		 T1K11	&	N/A 	&	0.77	&	0.60	&	0.99	&	N/A 	&	N/A\\
	\end{tabular}
	\caption{\label{table1} Measured eigenfrequencies of the six dominant oscillation modes in the 12 NS models, including the fundamental quasi-radial $(\ell=0)$ mode $F$ and its first overtone $H_1$, the fundamental quadrupole $(\ell=2)$ mode $^2f$ and its first overtone $^2p_1$, as well as the fundamental hexadecapole $(\ell=4)$ mode $^4f$ and its first overtone $^4p_1$, all predominantly excited under the perturbation with the corresponding $l$ index.
	All eigenfrequencies are in kHz and rounded off to two decimal places.
	The undetermined eigenfrequencies in specific models are denoted by `N/A'.
	The missing eigenfrequencies in the column of $F$ mode are due to unsatisfactory data quality in \texttt{Gmunu} simulations of T1K8 and T1K11 models under $\ell=0$ perturbation, while the missing eigenfrequencies in the columns of $^4f$ and $^4p_1$ modes are due to the hexadecapole $(\ell=4)$ modes being masked by the quadrupole $(\ell=2)$ modes in \texttt{Gmunu} simulations of the most magnetized models under $\ell=4$ perturbation.
	}
\end{table}

\begin{table}[ht]
	\centering
	\begin{tabular}{cccccccc}
		 Model & $\rho_{\textrm{c}}$ & $M$ & $R_{\textrm circ}$ & $r_{\textrm{e}}$ & $r_p/r_{\textrm e}$ & $\mathscr{H}/\mathscr{W}$ & $B_{\textrm{max}}$\\
		 & ($10^{14}$ g cm$^{-3}$) & ($M_\odot$) & (km) & (km) & & & ($10^{17}$ G)\\
		\hline
		 REF & 8.56 & 1.55 & 14.25 & 11.85 & 1.00 & 0.00 & 0.00\\
		 T1K1 & 8.56 & 1.55 & 14.25 & 11.85 & 1.00 & $3.97\times10^{-6}$ & $3.45\times10^{-2}$\\
		 T1K2 & 8.56 & 1.55 & 14.25 & 11.85 & 1.00 & $1.58\times10^{-5}$ & $6.89\times10^{-2}$\\
		 T1K3 & 8.57 & 1.55 & 14.25 & 11.85 & 1.00 & $3.95\times10^{-4}$ & $3.44\times10^{-1}$\\
		 T1K4 & 8.63 & 1.55 & 14.32 & 11.92 & 1.01 & $6.21\times10^{-3}$ & 1.36\\
		 T1K5 & 8.81 & 1.56 & 14.54 & 12.15 & 1.02 & $2.35\times10^{-2}$ & 2.63\\
		 T1K6 & 9.10 & 1.58 & 16.79 & 14.43 & 1.09 & $1.23\times10^{-1}$ & 5.52\\
		 T1K7 & 8.81 & 1.59 & 18.55 & 16.21 & 1.12 & $1.69\times10^{-1}$ & 6.01\\
		 T1K8 & 8.27 & 1.60 & 20.97 & 18.64 & 1.15 & $2.14\times10^{-1}$ & 6.14\\
		 T1K9 & 7.53 & 1.61 & 24.28 & 21.97 & 1.17 & $2.58\times10^{-1}$ & 5.96\\
		 T1K10 & 6.64 & 1.62 & 28.92 & 26.62 & 1.21 & $3.02\times10^{-1}$ & 5.53\\
		 T1K11 & 5.69 & 1.63 & 35.48 & 33.19 & 1.24 & $3.44\times10^{-1}$ & 4.93\\
	\end{tabular}
	\caption{\label{table2} Stellar properties of the 12 equilibrium models constructed by the \texttt{XNS} code.
	All numerical values are rounded off to two decimal places.
	$\rho_{\textrm{c}}$ is the central density, $M$ is the gravitational mass, $R_{\textrm circ}$ is the circumferential radius, $r_{\textrm{e}}$ is the equatorial radius, $r_p/r_{\textrm e}$ is the ratio of polar radius $r_p$ to equatorial radius $r_{\textrm{e}}$ (a purely toroidal field elongates the NS along the z-axis), $\mathscr{H}/\mathscr{W}$ is the ratio of total magnetic energy $\mathscr{H}$ to total binding energy $\mathscr{W}$, and $B_{\textrm{max}}$ is the maximum field strength achievable inside the star.
	These quantities are defined according to Ref.~\cite{2014MNRAS.439.3541P}.
	All the 12 models share the same rest mass $M_{\textrm{0}}=1.68$ M$_\odot$, polytropic constant $K=1.6\times10^5$ cm$^5$ g$^{-1}$ s$^{-2}$, and adiabatic index $\gamma=2$.
	The 11 magnetized models also share the same toroidal magnetization index $m=1$.}
\end{table}

\clearpage

\subsection*{\label{sec:cowling}Supplementary Note 1: Comparison to the Cowling approximation}
It is known that the Cowling approximation (i.e. fixed spacetime) overestimates the oscillation mode frequencies of NSs up to a factor of $\sim$ 2. Simulations with dynamical spacetime are essential for more accurate results. In this supplementary information, we show that the Cowling approximation is indeed inadequate under the circumstances described in this work. The corresponding visualizations of eigenfunctions of the excited oscillation modes are shown in Supplementary Figure~\ref{figs1}. Supplementary Table~\ref{tables1} summarizes the measured eigenfrequencies of the six modes in the 12 different NSs with the undetermined eigenfrequencies due to unsatisfactory data quality denoted by `N/A'. We plot in Supplementary Figure~\ref{figs2} the eigenfrequencies $f_{\textrm{eig}}$ of the six modes with and without Cowling approximation as functions of the magnetic to binding energy ratio $\mathscr{H}/\mathscr{W}$ of the NS model. We further summarize the relative differences in $f_{\textrm{eig}}$ with the Cowling approximation with respect to those without the approximation in Supplementary Table \ref{tables2} to better illustrate the discrepancies.  

We found a similar qualitative relation between $f_{\textrm{eig}}$ of the six modes and $\mathscr{H}/\mathscr{W}$ for both simulations with and without the Cowling approximation. However, a clear quantitative difference in the values of $f_{\textrm{eig}}$ is observed. The largest discrepancy is found in the $F$-mode, where the eigenfrequencies in all NS models are overestimated by a factor of $\sim$ 2 under the Cowling approximation. The discrepancies are also significant for $H_1$-mode with relative differences of $\gtrsim$ 20 per cent in most cases and even up to $\sim$ 65 per cent in the NS model with the strongest magnetic field strength. The discrepancies are less severe for higher-order modes (i.e. $\ell=2$ and $\ell=4$ modes). The relative differences are less than 20 per cent for $^2f$ and $^2p_1$ modes, while they are less than 10 per cent for $^4f$ and $^4p_1$ modes.

Hence, imposing the Cowling approximation in our simulations also results in overestimating the correct eigenfrequency up to a factor of $\sim$ 2. Dynamical spacetime is still necessary for magnetized NS simulations to obtain more realistic oscillation mode frequencies, especially for the lower-order modes.

\newpage

\begin{table}[ht]
	\centering
	\begin{tabular}{cccccccc}
		 Model & $F$ & $H_1$ & $^2f$ & $^2p_1$ & $^4f$ & $^4p_1$\\
		\hline
		 REF	&	2.70	&	4.61	&	N/A	&	4.22	&	2.63	&   5.26\\
		 T1K1	&	2.70	&	4.61	&	N/A	&	4.22	&	2.63    &   5.28\\
		 T1K2	&	2.70	&	4.61	&	N/A	&	4.22	&	2.64    &   5.26\\
		 T1K3	&	2.70	&	4.62	&	N/A	&	4.23	&	2.61	&	5.26\\
		 T1K4	&	2.70	&	4.51	&	1.90	&	4.10	&	2.60	&	5.10\\
		 T1K5	&	2.60	&	4.42	&	1.91	&	4.02	&	2.51	&	5.01\\
		 T1K6	&	2.10	&	3.63	&	1.60	&	3.30	&	2.10	&	4.14\\
		 T1K7	&	1.80	&	3.20	&	1.40	&	2.83	&	1.91	&	3.60\\
		 T1K8	&	1.50     &	2.70	&	1.20	&	2.40	&	1.60	&	3.01\\
		 T1K9	&	1.20	&	2.20	&	0.99	&	1.97	&	1.31     &	2.40\\
		 T1K10	&	N/A	&	1.70	&	0.78	&	1.51	&	0.99	    &	1.89\\
		 T1K11	&	0.68 	&	1.27	&	0.59	&	N/A	&	N/A 	&	1.40\\
	\end{tabular}
	\caption{\label{tables1} Measured eigenfrequencies of the six dominant oscillation modes \textbf{with the Cowling approximation} in the 12 NS models, including the fundamental quasi-radial $(\ell=0)$ mode $F$ and its first overtone $H_1$, the fundamental quadrupole $(\ell=2)$ mode $^2f$ and its first overtone $^2p_1$, as well as the fundamental hexadecapole $(\ell=4)$ mode $^4f$ and its first overtone $^4p_1$, all predominantly excited under the perturbation with the corresponding $l$ index.
	All eigenfrequencies are in kHz and rounded off to two decimal places.
	The undetermined eigenfrequencies in specific models are denoted by `N/A'. The missing eigenfrequencies are due to unsatisfactory data quality in \texttt{Gmunu} simulations under the perturbations.
	}
\end{table}

\begin{table}[ht]
	\centering
	\begin{tabular}{cccccccc}
		 Model & $F$ & $H_1$ & $^2f$ & $^2p_1$ & $^4f$ & $^4p_1$\\
		\hline
		 REF	&	+104.55	&	+16.71	&	N/A	&	+12.53	&	+5.20	&   +5.20\\
		 T1K1	&	+104.55	&	+16.71	&	N/A	&	+12.53	&	+5.20   &   +5.60\\
		 T1K2	&	+104.55	&	+16.71	&	N/A	&	+12.53	&	+5.60   &   +4.99\\
		 T1K3	&	+104.55	&	+17.56	&	N/A	&	+11.90	&	+3.98	&	+4.99\\
		 T1K4	&	+106.11	&	+15.64	&	+16.56	&	+10.81	&	+4.00	&	+3.87\\
		 T1K5	&	+100.00	&	+16.01	&	+18.63	&	+11.36	&	+4.58	&	+4.38\\
		 T1K6	&	+107.92	&	+24.74	&	+7.38	&	+13.79	&	+0.96	&	+3.76\\
		 T1K7	&	+100.00	&	+28.00	&	+6.06	&	+13.20	&	+6.11	&	+3.45\\
		 T1K8	&	N/A    &	+28.57	&	+1.69	&	+14.29	&	+7.38	&	+3.44\\
		 T1K9	&	+100.00	&	+29.41	&	+1.02	&	+15.88	&	N/A     &	0.00\\
		 T1K10	&	N/A	&	+27.82	&	0.00	&	+16.15	&	N/A	    &	+2.72\\
		 T1K11	&	N/A	&	+64.94	&	-1.67	&	N/A	&	N/A 	&	N/A\\
	\end{tabular}
	\caption{\label{tables2} Relative difference in per cent in the frequencies of the six dominant oscillation modes \textbf{with the Cowling approximation} with respect to those without the approximation. 
	The models containing undetermined eigenfrequencies are denoted by `N/A'.
	All relative differences are rounded off to two decimal places.
	}
\end{table}

\begin{figure}[ht]
	\centering
	\includegraphics[width=\linewidth]{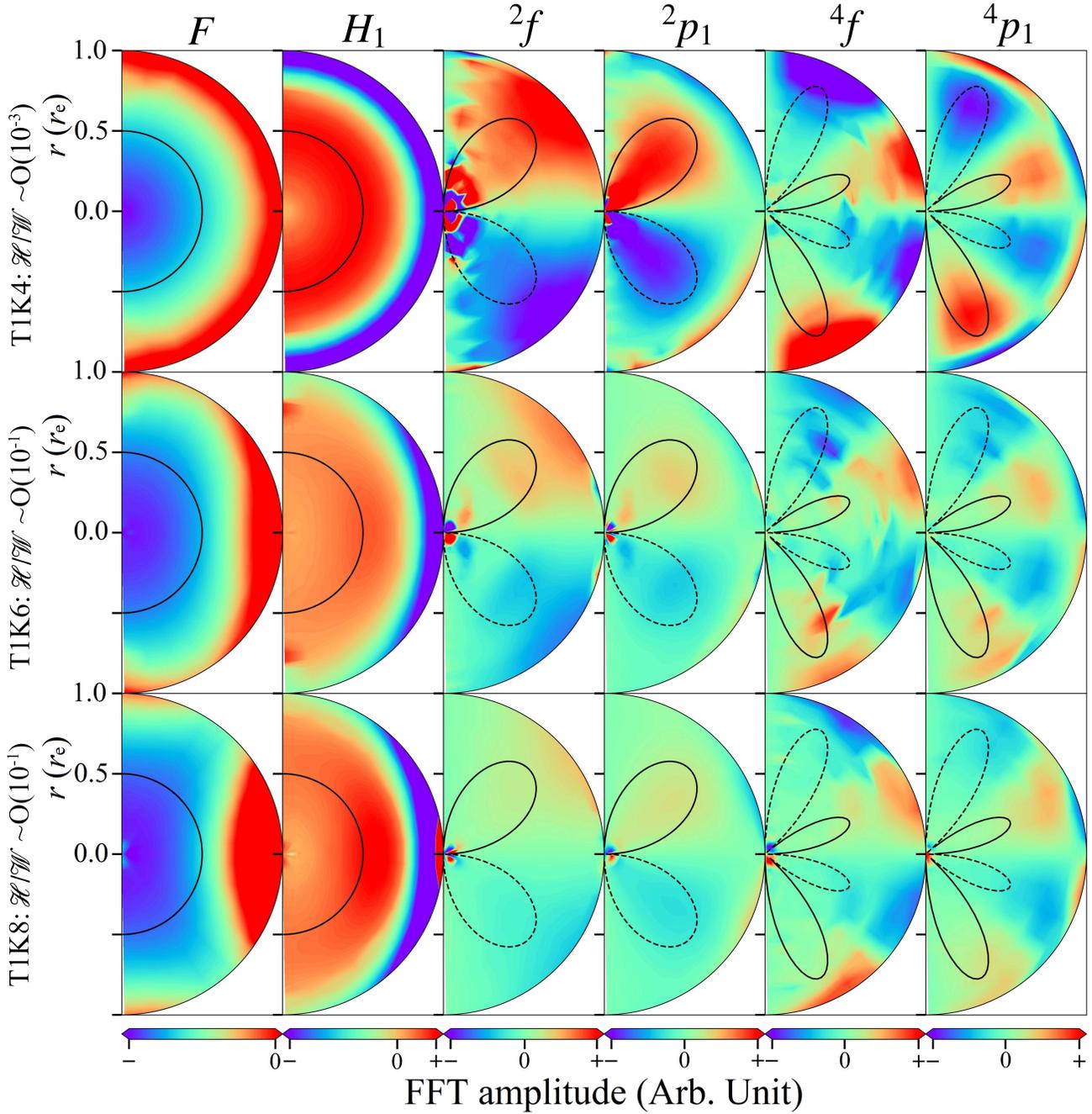}
	\caption{\label{figs1} Visualizations of eigenfunctions of the six dominant oscillation modes \textbf{with the Cowling approximation} using the data of 3 equilibrium models (T1K4, T1K6 and T1K8).
	The fundamental quasi-radial $(\ell=0)$ mode $F$ and its first overtone $H_1$ are predominantly excited under $\ell=0$ perturbation; the fundamental quadrupole $(\ell=2)$ mode $^2f$ and its first overtone $^2p_1$ are predominantly excited under $\ell=2$ perturbation; the fundamental hexadecapole $(\ell=4)$ mode $^4f$ and its first overtone $^4p_1$ are predominantly excited under $\ell=4$ perturbation.
	Each polar color plot shows the spatial map of FFT amplitude at the eigenfrequency of the mode, where the radial axis is normalized to the equatorial radius $r_{\textrm e}$ of each model.
	On top of each color plot, there is a polar line plot visualizing the $\theta$-part of the spherical harmonic in the corresponding perturbation function, where the distance from the origin to the line measures the magnitude of the spherical harmonic in that $\theta$-direction, while the solid and dotted portions represent the positive and negative parts of the spherical harmonic respectively.
	Each line plot is scaled arbitrarily for clearer illustration.
	It can be seen that the eigenfunctions of the higher-order quadrupole $(\ell=2)$ and hexadecapole $(\ell=4)$ modes have more nodes in the $\theta$-direction compared to the quasi-radial $(\ell=0)$ modes, while the eigenfunction of each first overtone has more nodes in the $r$-direction compared to its fundamental mode.
	Furthermore, each eigenfunction qualitatively agrees with the spherical harmonic in the corresponding perturbation function.}
\end{figure}

\begin{figure}[ht]
	\centering
	\includegraphics[height=0.8\textheight]{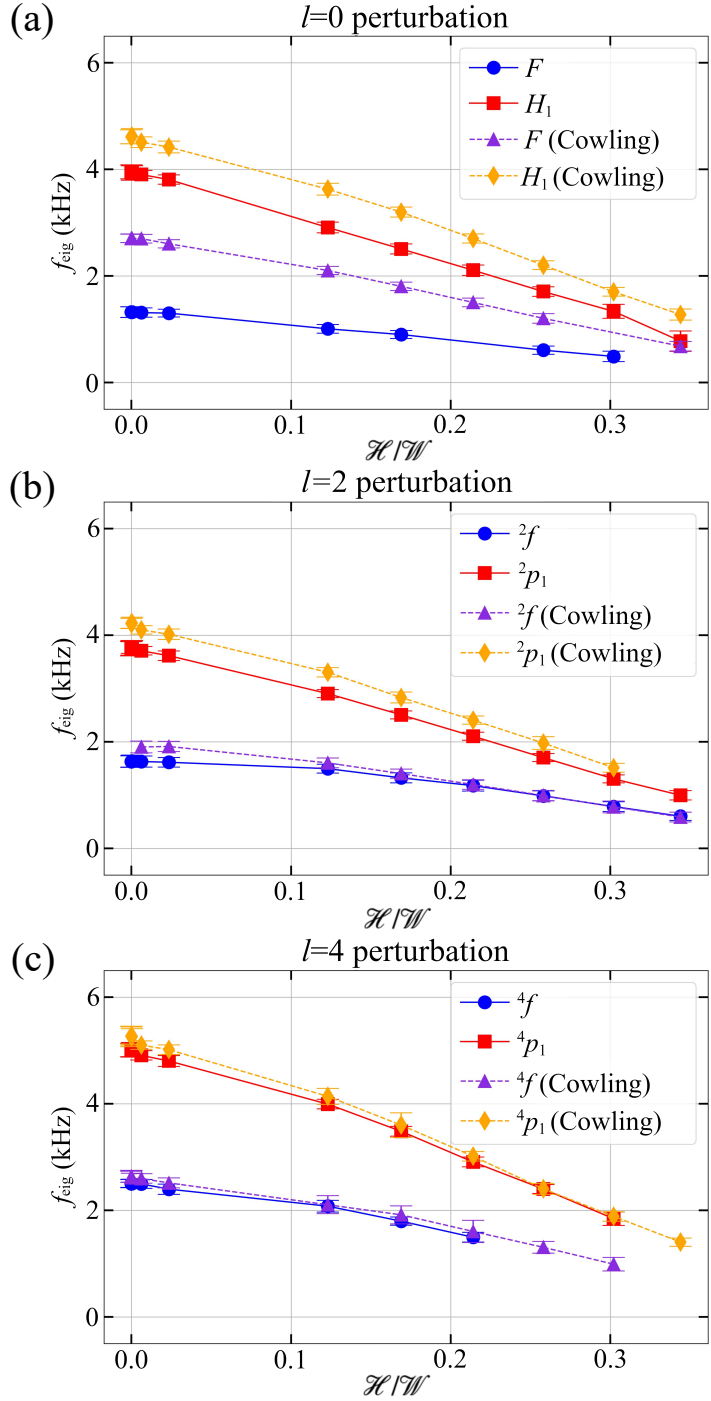}
	\caption{\label{figs2} Comparison of eigenfrequencies $f_{\textrm{eig}}$ of the excited oscillation modes \textbf{with and without the Cowling approximation} against the magnetic to binding energy ratio $\mathscr{H}/\mathscr{W}$ of the NS model, if $\ell=0$ (a), $\ell=2$ (b), and $\ell=4$ (c) perturbations are applied respectively.
	}
\end{figure}

\clearpage

\end{document}